\documentclass[sigconf,authordraft, review=False]{acmart}

\usepackage{float}
\usepackage{subcaption}
\usepackage{graphicx}
\usepackage{makecell}
\usepackage{multirow}
\usepackage{enumitem}
\usepackage[flushleft]{threeparttable}

\AtBeginDocument{%
  }

\begin{document}

\title{User Authentication and Identity Inconsistency Detection via Mouse-trajectory Similarity Measurement}

\author{Rui Jin}
\email{jinrui1@mail.ustc.edu.cn}
\affiliation{%
  \institution{University of Science and Technology of China}
  \city{Hefei}
  \state{Anhui}
  \country{China}
}

\author{Pengyuan Zhou}
\email{pyzhou@ustc.edu.cn}
\affiliation{%
  \institution{University of Science and Technology of China}
  \city{Hefei}
  \state{Anhui}
  \country{China}
}

\author{Yong Liao}
\email{yliao@ustc.edu.cn}
\affiliation{%
  \institution{University of Science and Technology of China}
  \city{Hefei}
  \state{Anhui}
  \country{China}
}

\renewcommand{\shortauthors}{Trovato et al.}





\begin{abstract}
Completely Automated Public Turing Test To Tell Computers and Humans Apart (CAPTCHA) is a type of challenge–response test widely used in authentication systems. A well-known challenge it faces is the CAPTCHA farm, where workers are hired to solve CAPTCHAs manually. In this work, we propose to tackle this challenge from a novel perspective, converting CAPTCHA farm detection to identity inconsistency detection, which essentially becomes an authentication process. Specifically, we develop a novel embedding model, which measures the similarity between mouse trajectories collected during the session and when registering/solving CAPTCHA, to authenticate and detect identity inconsistency. Moreover, unlike most existing works that employ a separate mouse movement classifier for each individual user, which brings in considerable costs when serving a large number of users, our model performs detection tasks using only one classifier for all users, significantly reducing the cost. Experiment results validate the superiority of our method over the state-of-the-art time series classification methods, achieving 94.3\% and 97.7\% of AUC in identity and authentication inconsistency detection, respectively.
\end{abstract}

\maketitle

\section{Introduction}
Handwriting analysis can be used as evidence in court. Signatures can be verified based on the kinematic theory of rapid human movements \cite{fischer2016signature}. Traditional biometrics such as fingerprint, iris, and face are only collected when they are needed for authentication or identification. Mouse movements, on the other hand, are captured whenever the user performs operations that require the participant of the cursor, thus making it possible to verify the user's identity throughout the session. Besides, since recording mouse behaviors does not require additional devices, the cost of deployment is relatively low.

In 2003, Everitt and McOwan first attempted to use the users' signatures drawn using a mouse to develop an authentication system \cite{everitt2003java}. Since then, researchers have attempted to realize mouse-based authentication using statistical methods, machine learning, and deep learning \cite{ahmed2007new,ahmed2011dynamic,zheng2016efficient,siddiqui2021continuous,fu2020rumba,shen2012user,chong2019user}. Surprisingly, studies that applied statistical analysis or manual feature engineering usually yield better results. It could be that such methods can extract information from typically 10 minutes to half an hour which is too long for deep learning approaches. However, such a long authentication time also leaves security risks as the attacker can accomplish its task before being detected. Another issue is the cost of training authentication classifiers because most existing studies trained an individual classifier for each registered user. This causes a rapid increase in processing power demands and an unbalanced dataset when scaling up. One-class SVM and multi-output deep learning have been introduced to solve this problem, but the results have not been encouraging. 


The usage of biometrics is not limited to authentication and identification. For example, mouse dynamics, keystroke behavior, and images of hands are used to detect automated bots \cite{chu2013blog,bera2021two}. Such tests, also known as Completely Automated Public Turing Test To Tell Computers and Humans Apart (CAPTCHA)~\cite{von2003captcha}, aim to recognize and block automated programs. CAPTCHAs are widely used in registration to avoid fake accounts in authentication systems. However, CAPTCHAs have made solving CAPTCHAs a profitable business~\cite{motoyama2010re}. The cost of such attacks can be as low as 3\$ per 1,000 CAPTCHAs\footnote{https://2captcha.com/}. More importantly, existing CAPTCHA can not defend farming since CAPTCHAs are designed to be easy for humans. This issue has been largely overlooked so far.

We notice that CAPTCHA farm detection and mouse-based authentication can share the same solution: measuring the similarity between two mouse trajectories. If the given mouse trajectories are matched by those recorded by the user it claims to be, it passes the authentication, and vice versa. Similarly, if the mouse trajectories recorded when solving the CAPTCHA have significant differences from those recorded later, we can predict that the CAPTCHA is solved by someone else. To build a model capable of calculating the similarity between two given mouse trajectories, we apply an embedding network to extract features from the input and train it using sample pairs. Furthermore, we propose to perform base sample selection and dynamic authentication to improve the model's performance. Our contributions include:
\begin{itemize}[leftmargin=*]
\item We propose a novel framework to measure the similarity of two mouse trajectories, for both user authentication and CAPTCHA farm detection.
\item We introduce the concept of embedding to extract features from mouse trajectories using the same model. This can lower the required number of classifiers significantly. 
\item We propose base sample selection and dynamic authentication, which utilize the diversity of mouse trajectories and can improve the performance of the model significantly.
\item The proposed model is tested on a hybrid dataset consisting of mouse movement records from 130 users in guided and unguided environments. The experiment results prove the effectiveness and robustness of our model.
\end{itemize}


\section{Related Works}
\label{related}
\subsection{Mouse-Based Authentication}
\label{related_authentication}
Mouse-based biometrics are usually studied for mouse-based authentication and CAPTCHA. Since mouse behaviors are commonly generated continuously after the user passes the authentication, the researchers found mouse-based biometrics suitable for continuous authentication to prevent hijacking. Ahmed and Issa analyzed 22 users' mouse behaviors during unguided sessions and obtained a false acceptance rate (FAR) of 2.4649\% and a false rejection rate (FRR) of 2.4614\% \cite{ahmed2007new}. Despite the high accuracy, it takes 13.55 minutes on average to detect an identity mismatch. Later, they proposed to apply sequential sampling to compute the confidence of accepting/rejecting dynamically over time \cite{ahmed2011dynamic}. FAR and FRR were close to 0\% using three-quarters of the time needed in the former study. However, this required a high demand for processing power. Thus, only 5 users were included in the experiment dataset. Zheng et al. used angle-based metrics and SVM for classification \cite{zheng2016efficient}. They achieved an equal error rate (EER) of 1.3\% with an average authentication time of 37.73 min. Researchers have also tried to expand the data-collecting environments. Siddiqui et al. recorded mouse dynamics while 10 users played the video game Minecraft and used Binary Random Forest classifiers to achieve an average accuracy of 92\% \cite{siddiqui2021continuous}.

An authentication time of as long as ten minutes to half an hour is too slow for traditional authentication tasks. RUMBA-Mouse, a CNN-RNN combined neural network model, yields a 3.16\% EER with an authentication delay of 6.11 seconds on average~\cite{fu2020rumba}.  Shen et al. collected mouse movements from 37 participants in a tightly controlled environment and achieved a FAR of 8.74\% and an FRR of 7.69\% with a corresponding authentication time of 11.8 seconds~\cite{shen2012user}. It is worth noticing that Shen et al. trained a one-class SVM classifier instead of a separate classifier for each user. Similarly, 2D-CNN has been applied to avoid training multiple classifiers: the model has multiple binary outputs, each representing the prediction of the authentication results for a different user, resulting in an EER of 10\% \cite{chong2019user}. Overall, mouse-based authentication requires further improvement in the accuracy, authentication time, robustness, and cost of training to be practical.

\subsection{Mouse-Based CAPTCHAs}
Most CAPTCHAs require the user to move their cursors. Chu et al. developed a client-side logger and a classifier based on the C4.5 algorithm to identify bots \cite{chu2018bot}. UNI-CAPTCHA, a CAPTCHA based on mouse, keyboard, and web behaviors, applied Hybrid biLSTM+Softmax to detect bots\cite{suzen2021uni}. Both studies have achieved an accuracy of over 99\%, proving that the analysis result of mouse behaviors can be an important reference for detecting bots. Acien et al. proposed to model the trajectories according to the Sigma-Lognormal model from the kinematic theory of rapid human movements and proved their BeCAPTCHA-Mouse can detect mouse trajectories generated by a GAN (Generative Adversarial Network) efficiently \cite{acien2022becaptcha}. More importantly, mouse-based CAPTCHAs have been widely deployed as a part of commercial CAPTCHAs such as GeeTest\footnote{https://www.geetest.com/}. However, existing methods can only distinguish bots from humans yet cannot detect CAPTCHA farm attacks.

\section{Methodology}
\label{method}
\subsection{Mouse Movement Embedding}
\subsubsection{Data Preprocessing}
\label{preprocessing section}
The first challenge we face to build a robust mouse movement embedding model is to preprocess mouse movement data collected in different environments. For example, solving CAPTCHAs can be viewed as a guided environment, but the user behaviors on the website are uncertain. 
Two preprocessing methods are widely adopted: extracting certain numbers of intentional mouse behaviors (such as clicks or movements), or segmenting the data with a fixed time window. 
The former method is combined with statistical analysis or feature engineering to process the data collected in unguided environments due to the variety of the length of mouse behaviors and lack of continuity. Studies using this method usually require longer authentication time.
On the other hand, the CAPTCHAs can generally be solved within one minute~\cite{weng2019towards}, which cannot meet the authentication time demands. The latter method, however, cannot be applied to data collected in unguided environments since the mouse might be stationary for an extended period. Besides, this might break the continuity of a mouse movement.

\begin{figure*}[h]
  \centering
  \includegraphics[width=0.6\textwidth]{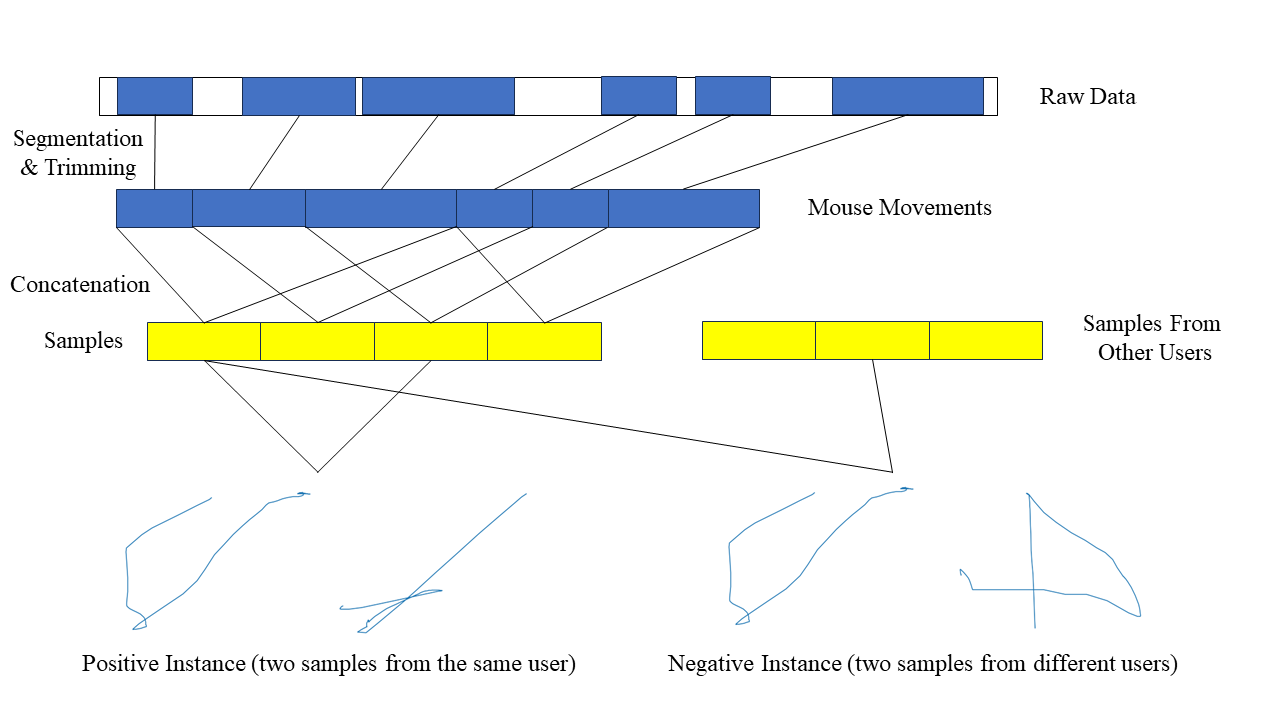}
  \caption{The preprocessing procedure. Meaningful mouse movements are cut from the raw data and then concatenated to generate samples, i.e., a time series that is long enough and can be used as one of the inputs. Two samples form instances. Our model measures the similarity between two samples in the instances and predicts whether they come from the same user.}
  \label{preprocessing}
\end{figure*}

To tackle this problem, we propose to perform segmentation, trimming, and concatenation on the moving state of the mouse. As shown in Fig.~\ref{preprocessing}, we first segment mouse movements to ensure the continuity of the mouse movements (blue blocks), then discard meaningless records and concatenate the segments to increase the information density. We cut the data where the state of the key is changed (pressed or released) or the gap between two consecutive timestamps is larger than 0.3 seconds. The segments consisting of less than 5 data points and those in which the cursor movement in the $x$ or the $y$ direction exceeds 5\% of the screen (white blocks) are considered to contain little mouse movement information and discarded. To provide samples with sufficient information for embedding and classification while keeping the continuity, we concatenate neighbor segments. Samples are generated using a sliding window with a certain maximum number of data points. All segments completely included in the window are concatenated to create a sample. The stride of the window is set to one segment. Since concatenating $(x,y)$ coordinate series directly results in unsmooth time series, we instead calculate the distance ($dx$ and $dy$) and the speed of the mouse ($\frac{dx}{dt}$ and $\frac{dy}{dt}$) along two directions between two data points. The sample length limit is set to 256. Thus, the input of our model is two matrices of shape $(256,4)$.

\subsubsection{Embedding}
Most existing studies train multiple classifiers, each performing a one-vs-rest binary classification. This method causes a heavy processing burden when the number of users increases. Shen et al. trained a legitimate-vs-illegitimate classifier, ignoring the difference between legitimate users \cite{shen2012user}. Chong et al. applied a 2D CNN with $N$ outputs \cite{chong2019user}. However, this is an ``All-in-one'' classifier that relies on the personal information stored in the network. Both studies attempting to use only one classifier failed to keep the performance comparable to other studies while doing so. We believe the reason is that they did not emphasize the difference between input data (their input is limited to one sample) because individual characteristics do influence the trained classifiers significantly.

\begin{figure}[h]
  \centering
  \includegraphics[width=0.98\columnwidth]{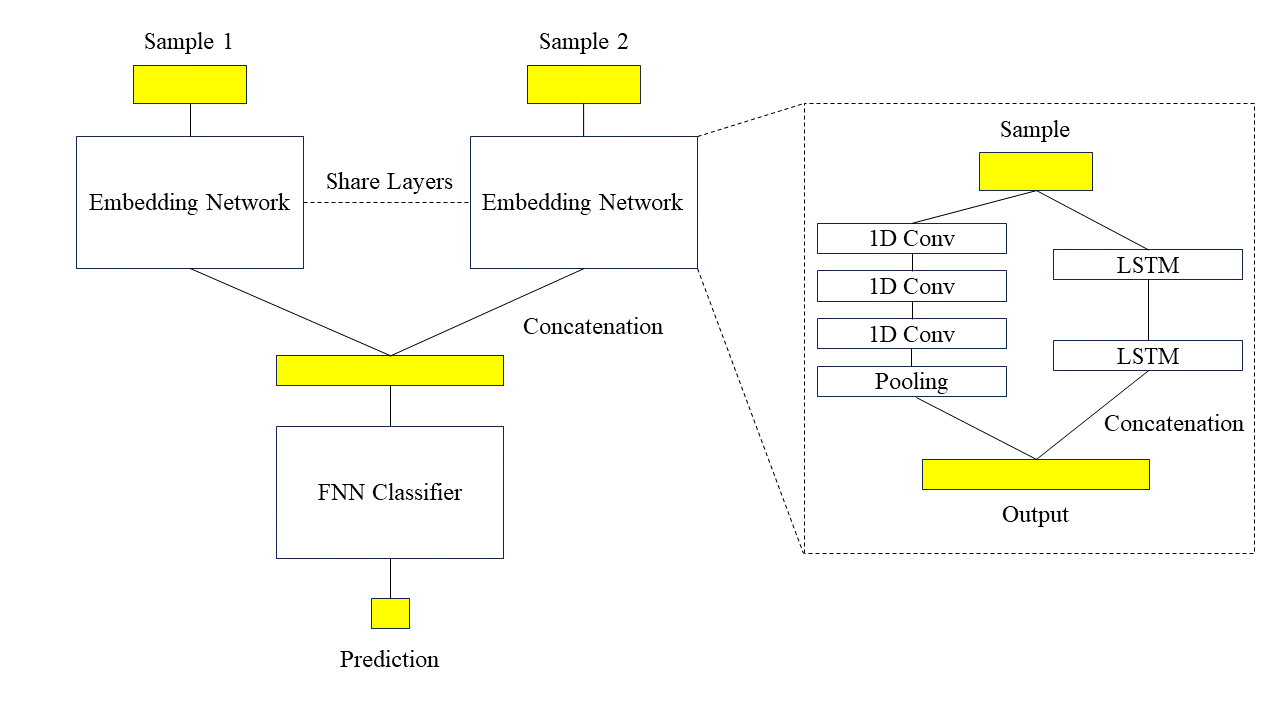}
  \caption{Classifier Framework}
  \label{classifier}
\end{figure}

The shared goal of authentication and CAPTCHA farm detection is to determine the similarity between the users. Widely deployed biometrics, such as fingerprints and faces, share a similar structure. When the user claims to be someone, the input data is mapped into a feature vector and compared with the corresponding feature vectors stored in the database. The authentication system can calculate their similarity and accept or decline the user's claim. We use this framework to avoid training individual classifiers for each user. The framework consists of an embedding network and a classifier. As shown in Fig.~\ref{classifier}, we input two samples each time, which are transformed into feature vectors using embedding networks with shared layers. The feature vectors are concatenated and passed to a classifier for similarity prediction.

The embedding network consists of two parallel networks: a three-layer 1D Convolutional Neural Network (1D CNN) and a two-layer Long short-term memory (LSTM) network. 2D CNN has been applied in numerous studies for image feature extraction. FaceNet, a famous model for face recognition, used 2D CNN for face embedding \cite{schroff2015facenet}. It is reasonable to perform convolutions on both dimensions of images, whereas only the convolutions along the time axis are proper for time series. Here, we stack three layers of 1D CNN with batch normalization and add a global average pooling layer to the end to reduce the dimensions. For sequence data, recurrent neural network (RNN) often performs well because it has hidden states that act as ``memory'' and can help store information about the processed part of the sequence. LSTM, a variation of RNN, can store long-term memory. Two layers of LSTM are used here to capture information from longer mouse trajectories. The classifier is a three-layer feedforward neural network (FNN) with dropouts to prevent overfitting. The network is trained with an Adam optimizer at a learning rate of 0.00001 using binary cross-entropy loss. The model is trained for 200 epochs for both authentication and CAPTCHA farm detection.

\begin{figure*}[h]
\begin{subfigure}{0.48\textwidth}
  \centering
  \includegraphics[width=3.5in]{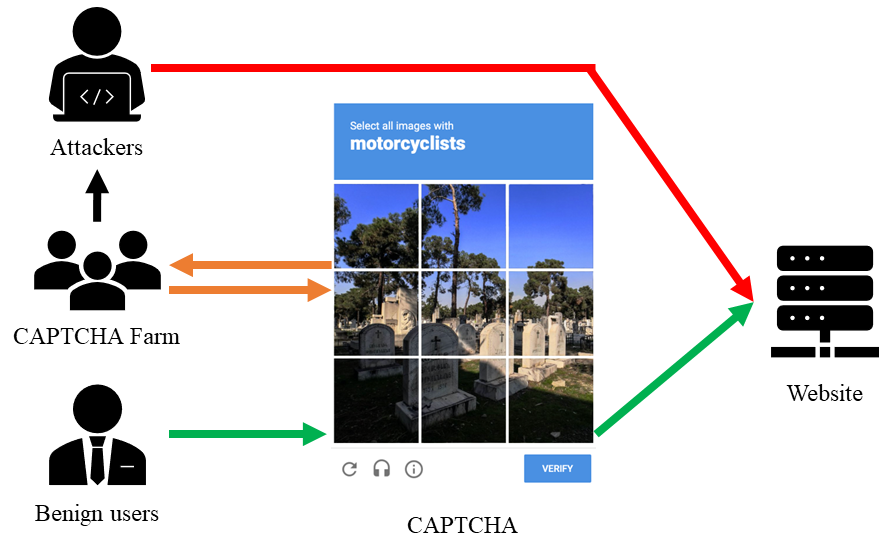}
  \caption{How attackers can breach CAPTCHAs using CAPTCHA farms. The image of CAPTCHA comes from Google reCAPTCHA\footnotemark.}
  \label{captcha_attack}
\end{subfigure}
\begin{subfigure}{0.48\textwidth}
  \centering
  \includegraphics[width=3.2in]{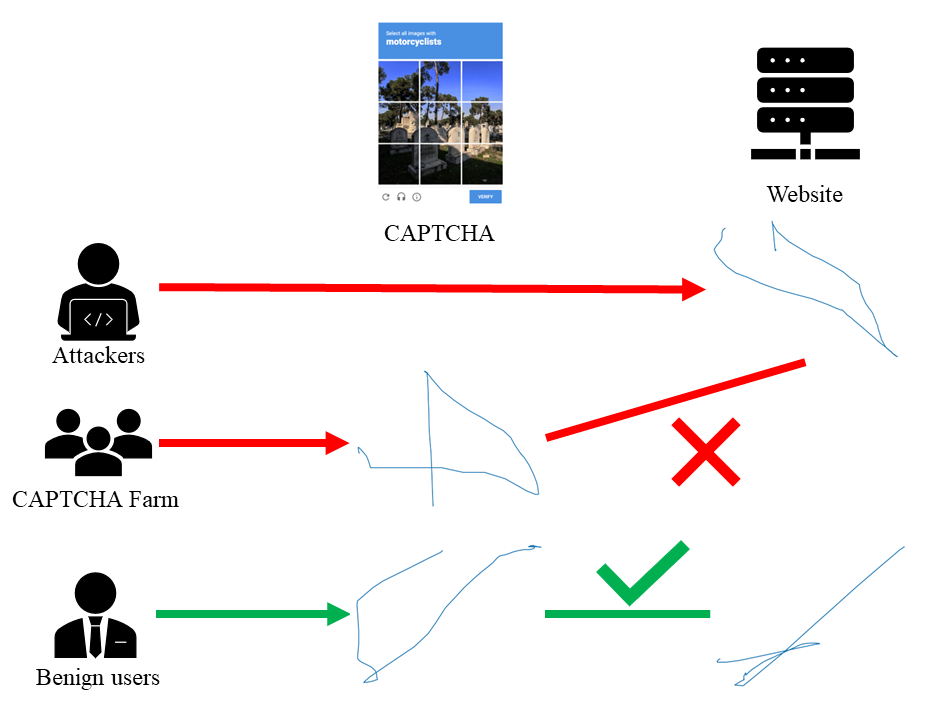}
  \caption{CAPTCHA Farms Detection.}
  \label{captcha_defense}
\end{subfigure}
\caption{The CAPTCHA farm attack and our solution.}
\end{figure*}
\footnotetext{https://developers.google.com/recaptcha}

\subsection{Identity Inconsistency Detection}
Today, most websites use CAPTCHAs from third-party CAPTCHA providers, which are usually implemented independently on the page. After solving the CAPTCHA, the collected data is sent to the CAPTCHA server. If the CAPTCHA provider believes the client is a human, a response token will be sent to the client. By verifying this token, the website knows whether the user is human. This verification procedure, however, has been breached by CAPTCHA-solving service retailers. They attack the CAPTCHA by pretending to be regular users and misappropriating tokens. As shown in Fig.~\ref{captcha_attack}, the attackers who perform malicious operations on the websites do not interact with the CAPTCHA providers directly. Instead, they send the necessary information to load the CAPTCHA to CAPTCHA farms. The CAPTCHA farm workers then solve the CAPTCHA and send response tokens they received from the CAPTCHA provider to the attackers. The attackers can use these tokens to pass the websites' verification.  

We found that such attacks are effective because the CAPTCHA provider and the website cannot verify the consistency of the user's identity. In this work, we propose to use biometrics to detect the attack. However, this is challenging because most modern CAPTCHAs require users to perform simple mouse movements like clicking or dragging, resulting in limited mouse behaviors collected for analysis.
Existing studies have developed continuous authentication systems using mouse-based biometrics to ensure identity consistency throughout the session. The reason they can not be applied to detect CAPTCHA farms is that they cannot distinguish unseen users: the workers and the attackers do not actively provide data for training. On the other hand, we propose to use mouse trajectory similarity to perform CAPTCHA farm detection as shown in Fig.~\ref{captcha_defense}. The mouse trajectory data when solving the CAPTCHA is stored until the session is closed. The website can record the user's mouse trajectory and check whether these two records are from different users. In other words, given two records, we aim to predict whether they come from the same user.

To train the classifier, we combine the generated samples to generate positive and negative instances. Assume that for the user with ID $i$, $n^{i}_{s}$ samples are generated. When generating positive instances, if we include all possible combinations, the number of positive instances for each user will be $\frac{n^{i}_{s}*(n^{i}_{s}-1)}{2}$. This could be too large for users with more data. We notice that neighbor sample pairs have large overlaps due to the sliding window. Therefore, such sample pairs can be easily recognized as positive and are far from practical. Thus, for a sample set $\{S^{i}_{0} , S^{i}_{1} , S^{i}_{2} , \ldots , S^{i}_{n^{i}_{s}-1}\}$, we generate positive instance set 
\begin{displaymath}
\{(S^{i}_{0},S^{i}_{\frac{n^{i}_{s}}{2}}),(S^{i}_{1},S^{i}_{1+\frac{n^{i}_{s}}{2}}),(S^{i}_{2},S^{i}_{2+\frac{n^{i}_{s}}{2}}),\ldots,(S^{i}_{\frac{n^{i}_{s}}{2}-1},S^{i}_{n^{i}_{s}-1})\}
\end{displaymath}
To generate a balanced dataset, for each positive instance, we combine one sample from it and another sample randomly chosen from other users to form a negative instance. After training the model with the generated dataset, it can be used for identity consistency verification.

\subsection{Mouse-Based Authentication}
Given mouse behavior data from $N$ different users, a record, and a target user ID, mouse-based authentication is to predict whether the record comes from the claimed target user. Due to the similarity between mouse-based authentication and identity inconsistency detection, we use a similar model structure and training process. 
There are two major differences between authentication and identity inconsistency detection. When performing identity inconsistency detection, both samples of the input sample pair come from unseen users, whereas one sample comes from the training data for authentication. Also, the authentication time is more flexible compared to identity inconsistency detection time, which is limited by CAPTCHA solving time.
We utilize these two differences to improve the classifier's performance further.

Given an input sample and a target user ID, we combine the sample with a base sample from the target user and pass it to the model to determine whether the user is the target user. Note that multiple base samples can be generated from the target user's training data. By validating their representativity using the validation set after training, we can pick the most representative samples for authentication. We encountered a combinatorial explosion issue when evaluating the base samples from the training set. We solved this issue by randomly selecting 20 samples from each user's validation set and randomly 20 samples from other users' validation set, then using them to evaluate the samples from the training set. We used the binary cross entropy loss function to evaluate the samples' representativity. The base sample selection can be represented as
\begin{displaymath}
\min_{j} \frac{1}{40} * \sum^{20}_{n=0} - (log(f(S^{i}_{j},S'^{i}_{n})) + log(1-f(S^{i}_{j},S'^{X}_{n})))
\end{displaymath}
where $f$ is the classifier, $S'^{i}$ is the shuffled validation set of user $i$, and $X$ has the discrete uniform distribution $f(x)=\frac{1}{N-1}$ for $x \in \{1,2,\ldots,N\} \setminus \{i\}$. $N$ represents the total number of users. 

We also found that due to the diversity of mouse trajectories, performing the classification multiple times using different samples benefits authentication accuracy. We use the half-overlapping sample in addition. Such expansion can be performed multiple times to trade authentication time for accuracy. Given sample $S^{i}_{j}$, let $e^{i}_{j}$ be the number of segments included in sample $S^{i}_{j}$, the expanded sample set can be expressed as:
\begin{displaymath}
\{S^{i}_{j}, E(S^{i}_{j}), E^{2}(S^{i}_{j}),\ldots, E^{samp_n-1}(S^{i}_{j})\}
\end{displaymath}
where $E(S^{i}_{j}) = S^{i}_{j+\frac{e^{i}_{j}}{2}}$ and $samp_n$ is an adjustable parameter representing the number of samples. The mean output of the classifications is calculated to provide a more reliable prediction.

\section{Experiments}
\label{experiment}
\subsection{Datasets}
We used two public datasets for evaluation. The SapiMouse dataset includes mouse trajectories from 120 users (92 male and 28 female) from the Sapientia University with ages between 18 and 53 years \cite{antal2021sapimouse}, in which a three-minute and a one-minute session were recorded for every user. This dataset was collected under a guided environment where users were required to accomplish as many mouse operations as possible in a little game\footnote{https://mousedynamicsdatalogger.netlify.app/}. The participants need to move the mouse on the randomly appeared icon and perform certain actions. Data collection was performed using a JavaScript web application running on the participants’ computers. This data acquisition protocol is very similar to CAPTCHA solving in terms of mouse trajectories. The Balabit Mouse Dynamics Challenge dataset consists of mouse trajectories from 10 users when they perform unspecified administrative tasks~\cite{Balabit}. A network monitoring device is set between the client and the remote computer that inspects all traffic as described by the RDP protocol. 5 to 7 sessions were recorded for each user. The session length varies from 40 minutes to 7 hours. Although some shorter sessions are provided as test data, we didn't use them since they are labeled as legal/illegal, and no identity information is provided. The SapiMouse dataset and the Balabit Mouse Dynamics Challenge dataset were combined to form the dataset used in our experiments so that data collected in guided and unguided environments were both learned. More details can be found in Table~\ref{datasets}. The Balabit dataset provides significantly more data for each user. 

Considering the ranges of coordinates depend on the resolution of the monitor, which varies between users and might cause an overly optimistic estimate, we normalized all coordinates before performing further preprocessing. We calculated the mouse movement time required for classification after preprocessing to exclude data with less meaningful mouse movement. On average, 18.5 seconds of effective mouse movement is used to generate a sample. 
\begin{table}[h]
\setlength{\tabcolsep}{4pt}
\centering
\caption{Datasets}
\label{datasets}
\begin{tabular}{|c|c|c|}
\hline
Dataset & SapiMouse & Balabit\\
\hline
No. users & 120 & 10\\
\hline
Environment & Guided & Unguided\\
\hline
\makecell[c]{Avg. time of mouse movements} & 156s & 9984s\\
\hline
\makecell[c]{Avg. time of data} & 247s & 63752s\\
\hline
\end{tabular}
\end{table}

\subsection{Evluation Metrics}
False Accept Rate (FAR) and False Reject Rate (FRR) are two essential metrics to evaluate a biometric system, measuring the possibility of allowing an attacker to pass and denying a legal user's access, respectively. Usually, a possibility instead of a deterministic result will be given by the classifier. The default passing threshold is set to 0.5 for FAR and FRR. The passing threshold can be adjusted to make the biometric system more secure (higher FRR and lower FAR) or less likely to trouble legal users (higher FAR and lower FRR). Thus, we sample the FRR and FAR at the passing threshold of $\{0,0.05,0.1,\ldots,1\}$ to draw the FRR-FAR curve to provide a more intuitive and comprehensive view. We also adopted the area under the curve (AUC) of the receiver operating characteristic (ROC) curve, which is a commonly used quantitative measurement of the overall performance of the classifier.

We also adopted two metrics concerning time, namely, the training time of the classifier and the authentication time. The training time reflects the training speed of a classifier. The authentication time is the length of mouse movement records required to accomplish authentication, which presents the convenience of the biometric system.
\begin{figure*}[h]
\caption{The FRR-FAR curves in identity inconsistency detection experiments.}
\begin{subfigure}{0.4\textwidth}
  \centering
  \includegraphics[width=2.8in]{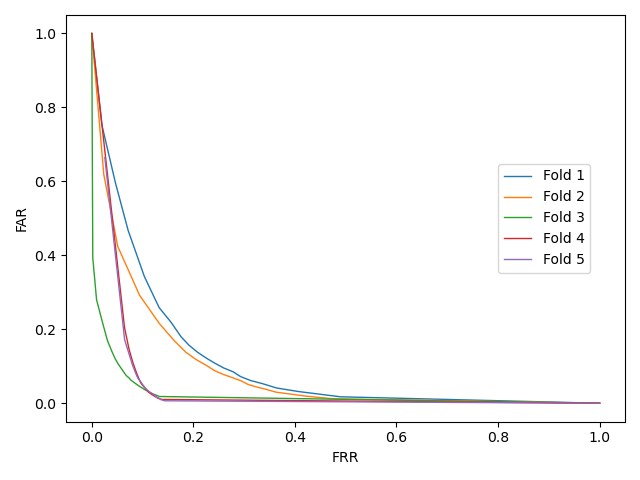}
  \caption{Our model's FRR-FAR curves of 5 folds (CNN+LSTM).}
  \label{FAR_FRR_fold}
\end{subfigure}
\begin{subfigure}{0.4\textwidth}
  \centering
  \includegraphics[width=2.8in]{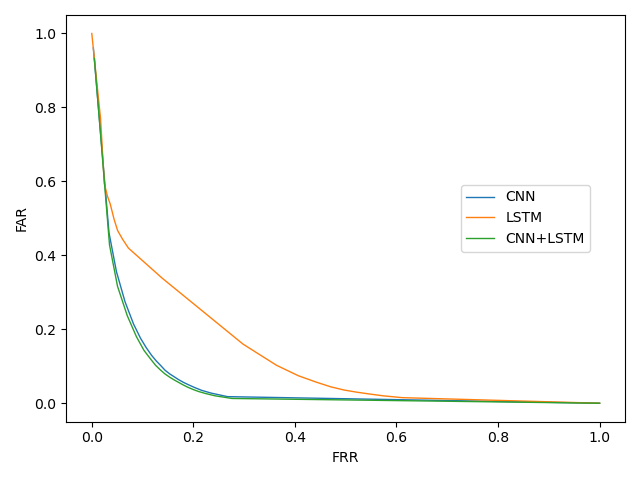}
  \caption{Mean FRR-FAR curves when using different embedding networks.}
  \label{FAR_FRR_embedding}
\end{subfigure}
\end{figure*}

\begin{table*}[h]
\centering
\caption{Comparation of MultiROCKET, HIVE-COTEv2, InceptionTime, and our model on identity inconsistency detection}
\label{identity}
\begin{threeparttable}
\begin{tabular}{|l|c|c|c|c|}
\hline
Method & AUC(\%) & False Accept Rate(\%) & False Reject Rate(\%) & Training Time(mins)\\
\hline
Hydra+MultiROCKET\cite{dempster2023hydra} & 87.7 & 19.0 & 19.5 & 798.7\\
\hline
HIVE-COTEv2\cite{middlehurst2021hive} & 85.4 & 21.2 & 18.0 & 505.2\\
\hline
InceptionTime\cite{ismail2020inceptiontime} & 93.6 & 21.2 & 9.6 & 527.9\\
\hline
Our Work\tnote{a} & 93.9 & 10.1 & 13.6 & 27.9\\
\hline
Our Work\tnote{b} & 85.6 & 16.0 & 29.8 & 56.2\\
\hline
Our Work\tnote{c} & 94.3 & 7.9 & 14.4 & 72.0\\
\hline
Our Work\tnote{d} & 94.3 & 18.0 & 8.8 & 72.0\\
\hline
Our Work\tnote{e} & 94.3 & 9.0 & 13.5 & 72.0\\
\hline
\end{tabular}
\begin{tablenotes}
\item[a] Excluding LSTM from the embedding network.
\item[b] Excluding CNN from the embedding network.
\item[c] Using CNN+LSTM in the embedding network.
\item[d] Using CNN+LSTM in the embedding network. Passing threshold was set to 0.75.
\item[e] Using CNN+LSTM in the embedding network. Passing threshold was set to 0.55.
\end{tablenotes}
\end{threeparttable}
\end{table*}

\subsection{Identity Inconsistency Detection}
\label{identity_experiment}
To simulate the threat scenario of a CAPTCHA farm attack, we randomly split the users in the dataset into training users and testing users without intersections, to keep the testing users unfamiliar with the classifier. Since few existing studies have attempted to address this attack, we applied three state-of-the-art time series classification models and used them as baselines: Hydra+MultiROCKET, HIVE-COTEv2, and InceptionTime. 
\begin{itemize}[leftmargin=*]
\item Hybrid Dictionary–Rocket Architecture (Hydra) combined dictionary method for time series classification and ROCKET \cite{dempster2020rocket}. Among tested variations, the combination of Hydra and MultiROCKET \cite{tan2022multirocket}, i.e., Hydra+MultiROCKET yields the best results \cite{dempster2023hydra}.
\item The Hierarchical Vote Collective of Transformation-based Ensembles (HIVE-COTE) is a heterogeneous meta ensemble for time series classification \cite{lines2018time}. HIVE-COTE 2.0 \cite{middlehurst2021hive}, as its upgraded variation, is an ensemble of four representative TSC models that use different approaches.
\item InceptionTime, a representative deep learning approach, is an ensemble of five deep learning networks based on Inception networks that are initialized randomly \cite{ismail2020inceptiontime}.
\end{itemize}
Besides comparing our model and the SOTA methods, we also take a look into the effect of CNN and LSTM in the embedding network.

We implemented MultiROCKET, HIVE-COTEv2, and InceptionTime using \texttt{aeon}, a Python toolkit for learning from time series\footnote{https://www.aeon-toolkit.org/}. For Hydra, we are using the multivariate version\footnote{https://github.com/angus924/hydra}. Since their training time is relatively long, we adjusted their training parameters and training datasets:
\begin{itemize}[leftmargin=*]
\item The number of training epochs of InceptionTime is set to 200 to align with our model. 
\item The scale of the training dataset for Hydra+MultiROCKET is limited to 20,000 instances randomly selected from the original training dataset. This limitation was set to control the training time near 720 minutes, which is 10 times our model's training time.
\item Similar to Hydra+MultiROCKET, the training dataset for HIVE-COTEv2 is limited to 10,000 instances. The training time limit was set to 720 minutes.
\end{itemize}
Although our limitation for Hydra+MultiROCKET and HIVE-COTEv2 
reduced the scale of their training dataset, their training time was 7 to 10 times ours even with the limitations, and removing the limitations would lead to unbearable training time. 
These models take one multivariate time series as input. To meet their requirements, we concatenate the two samples in each instance along the variates direction, leading to an input including $dx$, $dy$, $\frac{dx}{dt}$, and $\frac{dy}{dt}$ for both samples.

We used 5-fold cross-validation to ensure every user was considered during model performance evaluations. Experiments were carried out on a machine with an Intel Xeon Gold 5220 CPU at 2.20 GHz and a NVIDIA Tesla V100 GPU. Fig.~\ref{FAR_FRR_fold} shows the FRR-FAR curves of our model (See Appendix A for more FRR-FAR curves of Hydra+MultiROCKET, HIVE-COTEv2, and InceptionTime). Although differences in performance between folds can be overserved, the results are encouraging overall and can benefit CAPTCHA farm detection. We present the user-wise AUC score (the AUC score when using each user as the attacker) in Appendix C. The averaged results are presented in Table~\ref{identity}. InceptionTime and our model outperform Hydra+MultiROCKET and HIVE-COTEv2. 
Our classifier achieved the highest accuracy and the lowest FAR with a significantly shorter training time. InceptionTime obtained the lowest FRR, which is 4.8\% lower than ours, but its FRR is 13.3\% higher than ours. It is worth mentioning that if we adjust the passing threshold of our proposed model to 0.75, we can obtain a FAR of 18.0\% and a FRR of 8.8\%, which is 3.2\% and 0.8\% lower than the FAR and FRR of the InceptionTime method. Compared to the state-of-the-art time series classification methods, our model achieved better performance within a shorter training time.

\begin{figure*}[h]
\caption{The effect of training sample selection and dynamic authentication.}
\label{improvement}
\begin{subfigure}{0.3\textwidth}
  \centering
  \includegraphics[width=2.4in]{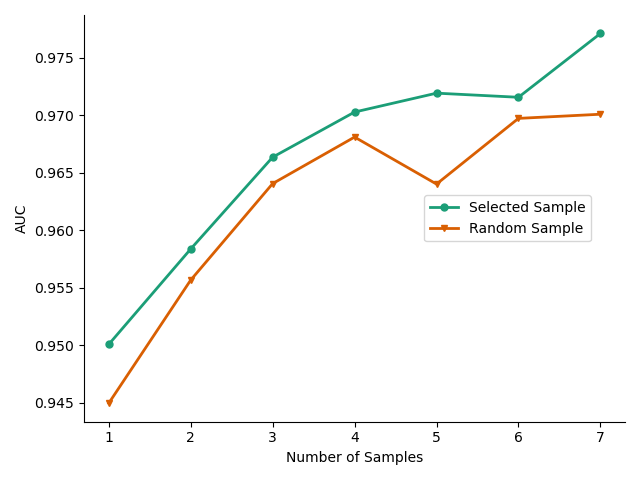}
  \caption{AUC}
\end{subfigure}
\begin{subfigure}{0.3\textwidth}
  \centering
  \includegraphics[width=2.4in]{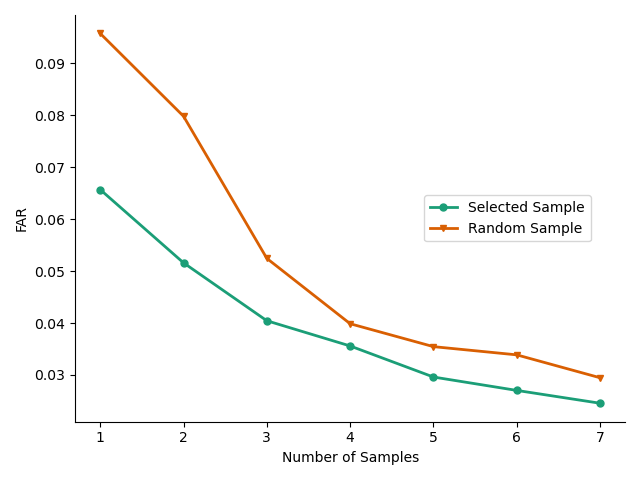}
  \caption{FAR}
\end{subfigure}
\begin{subfigure}{0.3\textwidth}
  \centering
  \includegraphics[width=2.4in]{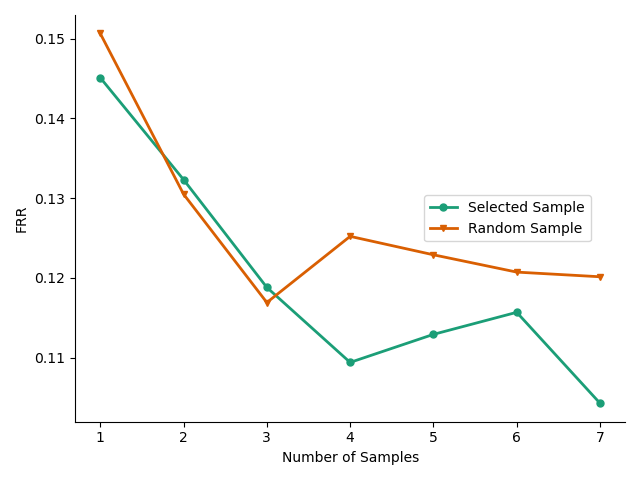}
  \caption{FRR}
\end{subfigure}
\end{figure*}

\begin{figure}[h]
  \centering
  \includegraphics[width=0.4\textwidth]{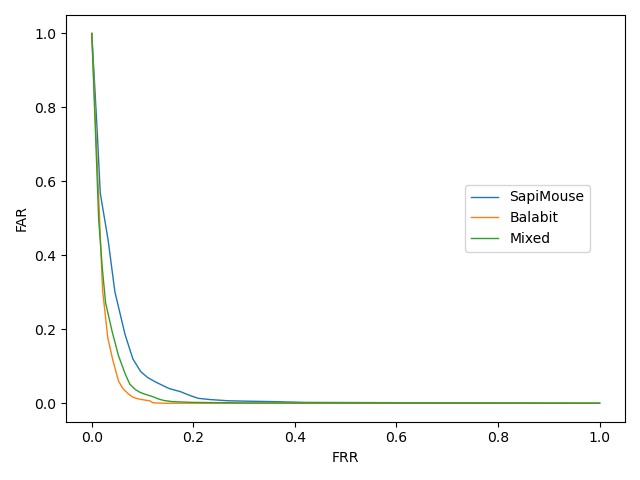}
  \caption{Authentication FRR-FAR curves using different registered user sets.}
  \label{FAR_FRR_2}
\end{figure}

\begin{figure}[h]
  \centering
  \includegraphics[width=0.4\textwidth]{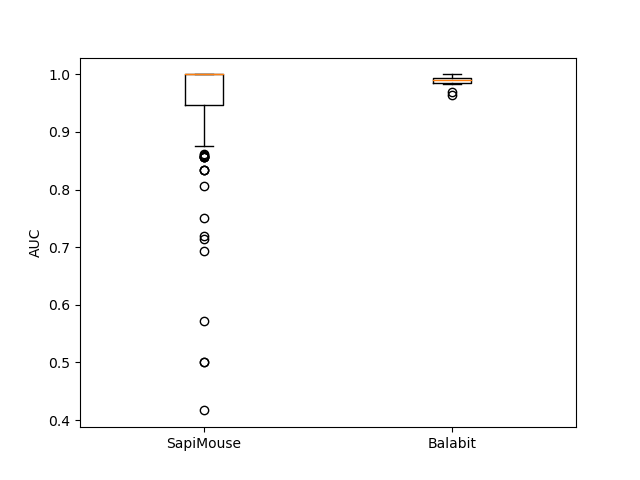}
  \caption{Distribution of AUC when viewing users in one dataset as registered users.}
  \label{box}
\end{figure}

\begin{table*}[h]
\centering
\caption{Comparation of the result of our authentication experiment and other representative studies}
\label{authentication}
\begin{threeparttable}
\begin{tabular}{|l|c|c|c|c|c|r|}
\hline
Study & \makecell{No. Registered Users} & \makecell{No. Attackers} & FAR(\%) & FRR(\%) & AUC(\%) & \makecell{No. Classifiers}\\
\hline
Fu et al.\cite{fu2020rumba} & 15 & 15 & 3.16 & 3.16 & 99.39 & 15 \\
\hline
Shen et al.\cite{shen2012user}\tnote{a} & 37 & 37 & 8.74 & 7.69 &  & 1\\
\hline
Chong et al.\cite{chong2019user} & 10 & 10 & 10 & 10 & 96 & 1\\
\hline
Antal et al.\cite{antal2021sapimouse} & 48 & 48 & 17 & 17 & 88 & 48\\
\hline
Our Work\tnote{b} & 130 & 127 & 5.2 & 7.5 & 97.7 & 1\\
\hline
Our Work\tnote{c} & 120 & 127 & 8.5 & 9.6 & 95.9 & 1\\
\hline
Our Work\tnote{d} & 10 & 127 & 3.1 & 7.3 & 98.2 & 1\\
\hline
\end{tabular}
\begin{tablenotes}
\item[a] This study did not provide the AUC score.
\item[b] Viewing all users as registered users. Passing threshold was set to 0.65.
\item[c] Viewing the users in the SapiMouse dataset as registered users. Passing threshold was set to 0.6.
\item[d] Viewing the users in the Balabit dataset as registered users. Passing threshold was set to 0.7.
\end{tablenotes}
\end{threeparttable}
\end{table*}

\noindent\textbf{Ablation.} We tested the effect of the CNN and LSTM by excluding them from the embedding network while keeping the other structures and parameters. Fig.~\ref{FAR_FRR_embedding} shows the mean FRR-FAR curves of 5 folds when using CNN, LSTM, and CNN+LSTM as the embedding network (see Appendix B for more detailed FRR-FAR curves of CNN and LSTM.). The performance of using LSTM alone is significantly lower than using CNN and CNN+LSTM. Although the FRR-FAR curve when using CNN alone is close to that for CNN+LSTM, CNN+LSTM outperforms CNN slightly overall, which is also confirmed by the AUC, FAR, and FRR summarized in Table~\ref{identity}. The AUC of CNN+LSTM is 0.4\% higher than CNN's AUC, and the FAR is 2.2\% lower despite the FRR being 0.8\% higher. If we adjust the passing threshold of CNN+LSTM to 0.55, a FAR of 9.0\% and a FRR of 13.5\% can be obtained, which is 1.1\% and 0.1\% lower than using CNN alone. However, regarding training time, using CNN alone takes only 27.9 minutes to train on average, which is roughly only one-third of the training time of CNN+LSTM. In summary, we found that both CNN and LSTM benefit the model's performance. Using CNN alone can reduce the training time at the cost of minor performance loss. CNN+LSTM yields better overall performance.

\subsection{Authentication}
Unlike the former experiment, since we had access to all users' data in this scenario, we used 80\% of each user's instances for training and the left 20\% for testing. No shuffling was performed before the train-test split to minimize the overlap between training and testing instances. We conducted two experiments to verify the effectiveness of the training sample selection and dynamic authentication and evaluate the performance of our model.

\subsubsection{Effect of base sample selection and dynamic authentication}
In the first experiment, we present how base sample selection and dynamic authentication can improve the performance of our model. For the baseline, we replace the selected base samples with a randomly selected sample from the user's training set. As shown in Fig.~\ref{improvement}, with base sample selection, higher AUC, lower FAR, and lower FRR were obtained. Improvement can also be observed as the number of samples increases, with or without the base sample selection. With 3 samples, i.e., doubled authentication time, the AUC, FAR, and FRR were improved by 2.1\%, 4.3\%, and 3.4\%, respectively. Although better performance might be observed with more samples, we stopped at 7 samples since this is the point where the authentication time increases by 3 times on average and grows over 1 minute. Compared to the poorest result obtained without base sample selection and dynamic authentication, the best result is 3.2\% higher in AUC, 7.1\% lower in FAR, and 4.6\% lower in FRR. This proves that base sample selection and dynamic authentication can improve our model's performance vastly. For the rest of the experiments, we used base sample selection and 7 samples if not otherwise mentioned.

\subsubsection{Authentication performance}
Next, we evaluate the performance of our model compared to other studies under 3 different settings: assuming as registered users all the users, only the ones in the SapiMouse dataset, and only the ones in the Balabit dataset. The reason is that despite mixing the dataset can showcase our model's generalization, the difference in data volume between the datasets varies the performance significantly, resulting in unavoidable bias. We consider all users as the attackers, i.e., samples from every user might be selected to generate a negative instance (three users were excluded when the number of samples was set to 7 since the volume of their data failed to meet the requirement of the dynamic authentication). As shown in Table~\ref{authentication}, the AUC score for all users considered as registered users is 97.7\%, 1.8\% higher than that when only SapiMouse dataset users were considered and 0.5\% lower than that when only Balabit dataset users were considered. As shown in Fig.~\ref{FAR_FRR_2}, the differences in FAR and FRR present a similar tendency, indicating that the model performs better on the Balabit dataset. Fig.~\ref{box} depicts the distribution of the AUC of each user. It can be observed that the AUC of users in the SapiMouse dataset is more distributed. We believe the reason is that for each user more training data can be extracted from the Balabit dataset than from the SapiMouse dataset. Please refer to Appendix C for detailed user-wise AUC scores. 
Table~\ref{authentication} also presents other representative studies that achieved mouse-based authentication. Note that we excluded those requiring longer than 10-minute continuous authentication process because such methods are too slow for traditional authentication tasks, as mentioned in Section ~\ref{related_authentication}. Our model achieved the second lowest FAR of 5.2\% and the second lowest FRR of 7.5\%. The FAR, FRR, and AUC are second only to one study~\cite{fu2020rumba}. Furthermore, we reduced the number of classifiers to 1. As far as we know, only two existing studies~\cite{shen2012user,chong2019user} managed to do so, and our model outperformed them. We reduced the FAR by 3.5\% while maintaining a lower FRR than one study ~\cite{shen2012user} and improved the AUC by 1.7\% compared to another~\cite{chong2019user}.
It is worth mentioning that by applying dynamic authentication, our model is more flexible in authentication time and can be further adjusted to meet different requirements.

\section{Conclusion}
Noticing the lack of studies concerning CAPTCHA farm detection and the need to reduce the number of classifiers for mouse-based authentication, we developed a novel model to kill two birds with one stone by converting CAPTCHA farm detection to identity inconsistency detection. Inspired by fingerprints and face embedding, we realized authentication and identity inconsistency detection with one model to measure the similarity of two given mouse trajectories. Additionally, we propose to utilize the diversity of mouse movement records to implement base sample selection and dynamic authentication, which can significantly improve the security and flexibility in authentication time. We tested our model on a mixed dataset consisting of 130 users. Our model obtained an AUC score of 94.3\% for identity inconsistency detection and 97.7\% for authentication. The experiment results prove that our model can achieve outstanding performance using only one classifier, and can detect inconsistency in the identity of unseen users, which can be a novel method to detect CAPTCHA farm attacks.


\bibliographystyle{ACM-Reference-Format}
\bibliography{main}

\clearpage
\appendix
\section{Hydra+MultiROCKET's, HIVE-COTEv2's, and InceptionTime's FRR-FAR curves}
\begin{figure}[h]
\begin{subfigure}{0.49\textwidth}
  \centering
  \includegraphics[width=3.2in]{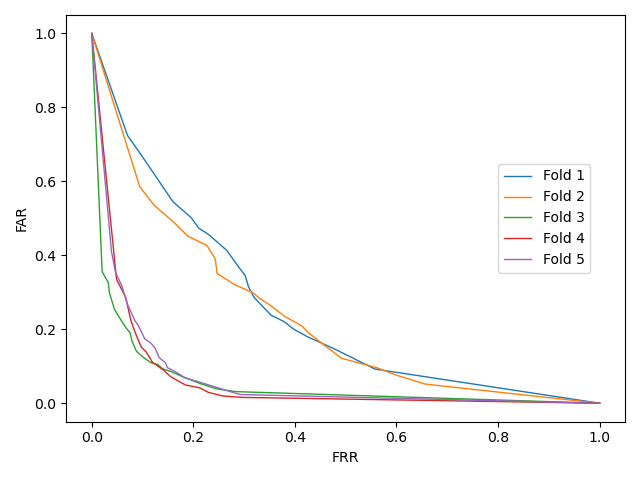}
  \caption{FRR-FAR curves of 5 folds when using Hydra+MultiROCKET.}
\end{subfigure}
\begin{subfigure}{0.49\textwidth}
  \centering
  \includegraphics[width=3.2in]{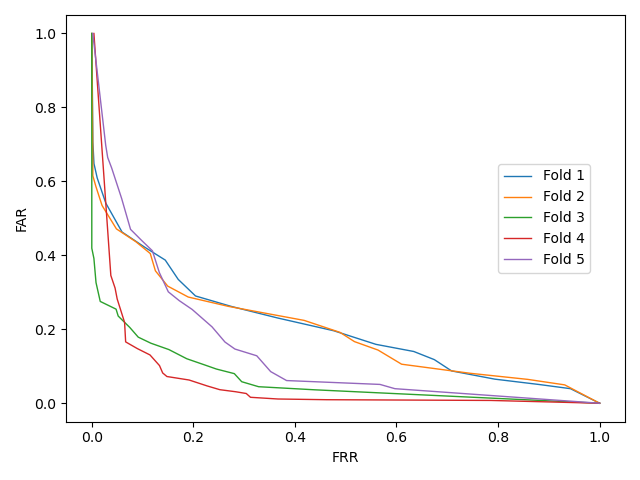}
  \caption{FRR-FAR curves of 5 folds when using HIVE-COTEv2.}
\end{subfigure}
\begin{subfigure}{0.49\textwidth}
  \centering
  \includegraphics[width=3.2in]{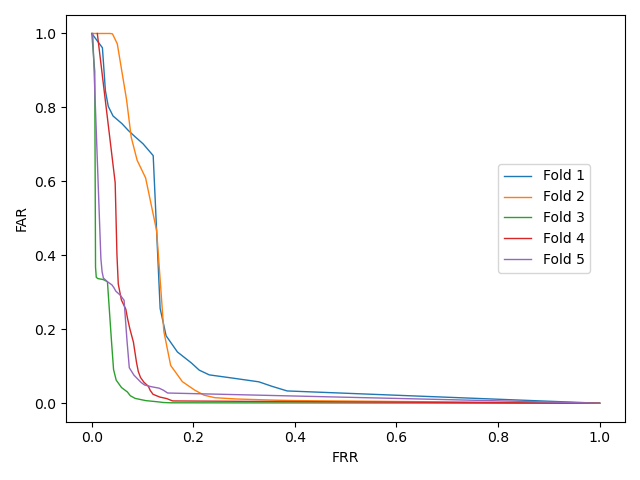}
  \caption{FRR-FAR curves of 5 folds when using InceptionTime.}
\end{subfigure}
\end{figure}

Here, we present the FRR-FAR curves of InceptionTime, HIVE-COTEv2, and Hydra+MultiROCKET when performing the 5-fold cross-validation we introduced in Section. ~\ref{identity_experiment}. Among these state-of-the-art time series classification methods, InceptionTime achieved significantly better performance. Hydra+MultiROCKET and HIVE-COTEv2 obtained less promising results.

\section{CNN's and LSTM's FRR-FAR curves}
\begin{figure}[h]
\begin{subfigure}{0.49\textwidth}
  \centering
  \includegraphics[width=3.2in]{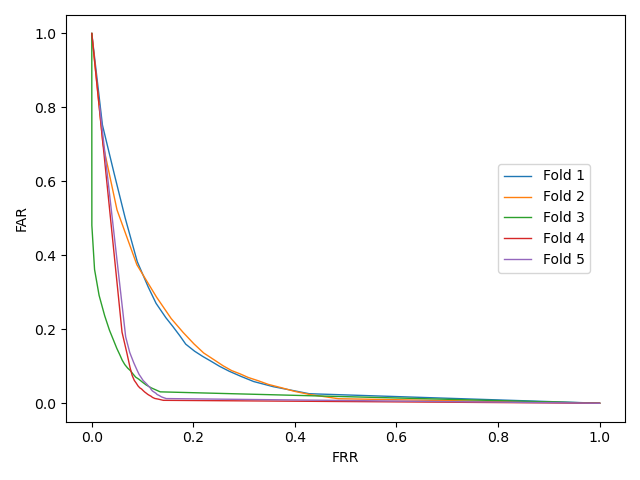}
  \caption{FRR-FAR curves of 5 folds when using CNN.}
\end{subfigure}
\begin{subfigure}{0.49\textwidth}
  \centering
  \includegraphics[width=3.2in]{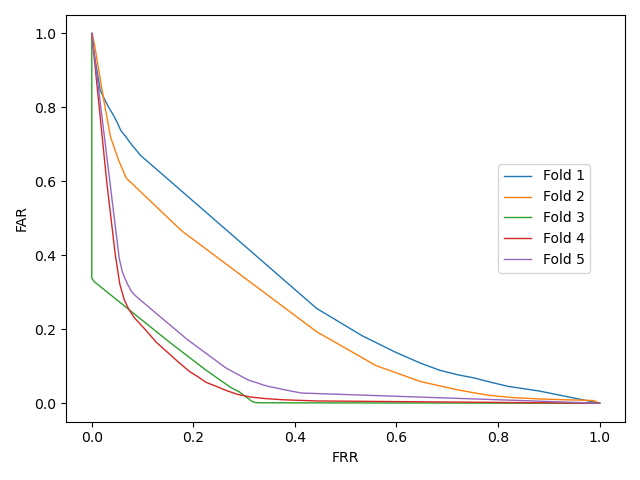}
  \caption{FRR-FAR curves of 5 folds when using LSTM.}
\end{subfigure}
\end{figure}

The preceding figures are the FRR-FAR curves when using CNN and LSTM alone as the embedding network (Section. ~\ref{identity_experiment}). The FRR-FAR curves when using CNN are obviously lower than those when using LSTM.

\clearpage
\section{User-wise AUC}
The following tables are the user-wise AUC score in the identity inconsistency detection and authentication experiment. User 1-120 come from the SapiMouse dataset. User 121-130 come from the Balabit dataset. Despite the mean AUC achieved in identity 
\begin{table}[H]
\centering
\caption{Identity Inconsistency Detection.}
\begin{tabular}{|c|c|c|c|c|c|}
\hline
User Id & AUC & User Id & AUC & User Id & AUC\\
\hline
1 & 85.0 & 46 & 72.4 & 91 & 89.9\\
\hline
2 & 85.4 & 47 & 91.8 & 92 & 73.6\\
\hline
3 & 88.5 & 48 & 84.9 & 93 & 85.7\\
\hline
4 & 58.7 & 49 & 82.3 & 94 & 88.0\\
\hline
5 & 86.8 & 50 & 88.7 & 95 & 99.3\\
\hline
6 & 94.3 & 51 & 92.7 & 96 & 92.3\\
\hline
7 & 83.6 & 52 & 73.7 & 97 & 98.9\\
\hline
8 & 76.5 & 53 & 88.3 & 98 & 98.2\\
\hline
9 & 93.0 & 54 & 98.5 & 99 & 79.3\\
\hline
10 & 85.8 & 55 & 71.0 & 100 & 94.3\\
\hline
11 & 98.1 & 56 & 99.0 & 101 & 77.4\\
\hline
12 & 87.3 & 57 & 98.2 & 102 & 86.6\\
\hline
13 & 87.3 & 58 & 96.4 & 103 & 89.2\\
\hline
14 & 88.4 & 59 & 91.2 & 104 & 83.2\\
\hline
15 & 91.5 & 60 & 60.3 & 105 & 91.4\\
\hline
16 & 78.9 & 61 & 85.6 & 106 & 88.6\\
\hline
17 & 82.0 & 62 & 85.3 & 107 & 95.7\\
\hline
18 & 77.1 & 63 & 84.5 & 108 & 88.4\\
\hline
19 & 99.5 & 64 & 94.4 & 109 & 69.2\\
\hline
20 & 99.0 & 65 & 91.7 & 110 & 90.2\\
\hline
21 & 98.2 & 66 & 73.4 & 111 & 94.1\\
\hline
22 & 77.5 & 67 & 89.0 & 112 & 84.4\\
\hline
23 & 87.1 & 68 & 83.0 & 113 & 87.0\\
\hline
24 & 94.8 & 69 & 92.5 & 114 & 99.1\\
\hline
25 & 96.1 & 70 & 81.8 & 115 & 83.8\\
\hline
26 & 97.8 & 71 & 82.0 & 116 & 99.5\\
\hline
27 & 75.2 & 72 & 79.8 & 117 & 99.2\\
\hline
28 & 91.2 & 73 & 98.9 & 118 & 79.2\\
\hline
29 & 85.9 & 74 & 93.9 & 119 & 84.6\\
\hline
30 & 90.7 & 75 & 96.0 & 120 & 100.0\\
\hline
31 & 86.1 & 76 & 98.1 & 121 & 90.8\\
\hline
32 & 82.5 & 77 & 75.3 & 122 & 97.6\\
\hline
33 & 92.2 & 78 & 94.0 & 123 & 96.7\\
\hline
34 & 89.7 & 79 & 92.7 & 124 & 96.3\\
\hline
35 & 83.4 & 80 & 94.1 & 125 & 100.0\\
\hline
36 & 79.6 & 81 & 92.9 & 126 & 91.8\\
\hline
37 & 78.5 & 82 & 89.7 & 127 & 98.1\\
\hline
38 & 94.7 & 83 & 82.5 & 128 & 96.3\\
\hline
39 & 73.5 & 84 & 100.0 & 129 & 95.5\\
\hline
40 & 98.8 & 85 & 95.9 & 130 & 96.3\\
\hline
41 & 95.0 & 86 & 90.8 & & \\
\hline
42 & 79.3 & 87 & 94.4 & & \\
\hline
43 & 95.5 & 88 & 87.6 & & \\
\hline
44 & 90.0 & 89 & 94.9 & & \\
\hline
45 & 77.9 & 90 & 93.4 & & \\
\hline
\end{tabular}
\end{table}

\noindent
inconsistency detection being lower than that achieved in authentication, the opposite conclusion can be reached if only the AUC of some specific users is considered. For example, the AUC of user 101 is 77.4\% for identity inconsistency detection and 41.7\% for authentication. This suggests that the limited training data from some users didn't benefit the classification but could be misleading instead.

\begin{table}[H]
\centering
\caption{Authentication. User 94, 97, and 98 were excluded.}
\begin{tabular}{|c|c|c|c|c|c|}
\hline
User Id & AUC & User Id & AUC & User Id & AUC\\
\hline
1 & 100.0 & 46 & 100.0 & 91 & 100.0\\
\hline
2 & 100.0 & 47 & 100.0 & 92 & 100.0\\
\hline
3 & 98.4 & 48 & 92.0 & 93 & 100.0\\
\hline
4 & 80.6 & 49 & 100.0 & 94 & \\
\hline
5 & 85.7 & 50 & 100.0 & 95 & 100.0\\
\hline
6 & 95.9 & 51 & 100.0 & 96 & 89.8\\
\hline
7 & 100.0 & 52 & 100.0 & 97 & \\
\hline
8 & 100.0 & 53 & 100.0 & 98 & \\
\hline
9 & 100.0 & 54 & 100.0 & 99 & 100.0\\
\hline
10 & 96.0 & 55 & 85.7 & 100 & 100.0\\
\hline
11 & 100.0 & 56 & 100.0 & 101 & 41.7\\
\hline
12 & 100.0 & 57 & 100.0 & 102 & 87.8\\
\hline
13 & 100.0 & 58 & 96.7 & 103 & 100.0\\
\hline
14 & 100.0 & 59 & 100.0 & 104 & 100.0\\
\hline
15 & 71.9 & 60 & 100.0 & 105 & 94.8\\
\hline
16 & 71.4 & 61 & 100.0 & 106 & 100.0\\
\hline
17 & 100.0 & 62 & 95.2 & 107 & 100.0\\
\hline
18 & 100.0 & 63 & 69.4 & 108 & 90.6\\
\hline
19 & 97.2 & 64 & 93.6 & 109 & 91.4\\
\hline
20 & 100.0 & 65 & 100.0 & 110 & 100.0\\
\hline
21 & 100.0 & 66 & 100.0 & 111 & 100.0\\
\hline
22 & 100.0 & 67 & 87.5 & 112 & 99.2\\
\hline
23 & 100.0 & 68 & 86.1 & 113 & 93.8\\
\hline
24 & 100.0 & 69 & 83.3 & 114 & 100.0\\
\hline
25 & 100.0 & 70 & 83.3 & 115 & 100.0\\
\hline
26 & 100.0 & 71 & 83.3 & 116 & 100.0\\
\hline
27 & 100.0 & 72 & 100.0 & 117 & 100.0\\
\hline
28 & 100.0 & 73 & 95.3 & 118 & 50.0\\
\hline
29 & 85.7 & 74 & 99.4 & 119 & 87.8\\
\hline
30 & 100.0 & 75 & 100.0 & 120 & 100.0\\
\hline
31 & 100.0 & 76 & 100.0 & 121 & 97.0\\
\hline
32 & 97.2 & 77 & 96.9 & 122 & 98.9\\
\hline
33 & 50.0 & 78 & 85.7 & 123 & 100.0\\
\hline
34 & 98.8 & 79 & 100.0 & 124 & 99.2\\
\hline
35 & 85.8 & 80 & 100.0 & 125 & 98.4\\
\hline
36 & 86.0 & 81 & 100.0 & 126 & 98.8\\
\hline
37 & 100.0 & 82 & 100.0 & 127 & 99.2\\
\hline
38 & 100.0 & 83 & 100.0 & 128 & 99.4\\
\hline
39 & 100.0 & 84 & 95.7 & 129 & 96.4\\
\hline
40 & 100.0 & 85 & 96.9 & 130 & 100.0\\
\hline
41 & 100.0 & 86 & 100.0 & & \\
\hline
42 & 100.0 & 87 & 75.0 & & \\
\hline
43 & 98.0 & 88 & 100.0 & & \\
\hline
44 & 90.6 & 89 & 100.0 & & \\
\hline
45 & 57.1 & 90 & 100.0 & & \\
\hline
\end{tabular}
\end{table}










\end{document}